# Agentic AI-Enabled Framework for Thermal Comfort and Building Energy Assessment in Tropical Urban Neighborhoods

*Po-Yen* Lai[1*], *Xinyu* Yang[1*], *Derrick* Low[1], *Huizhe* Liu[1], and *Jian Cheng* Wong[1]

[1]Institute of High Performance Computing (IHPC), Agency for Science Technology and Research (A*STAR), 1 Fusionopolis Way, #16-16 Connexis, Singapore 138632, Republic of Singapore.

**Abstract.** In response to the urban heat island effects and building energy demands in Singapore, this study proposes an agentic AI-enabled reasoning framework that integrates large language models (LLMs) with lightweight physics-based models. Through prompt customization, the LLMs interpret urban design tasks, extract relevant policies, and activate appropriate physics-based models for evaluation, forming a closed-loop reasoning-action process. These lightweight physics-based models leverage core thermal and airflow principles, streamlining conventional models to reduce computational time while predicting microclimate variables, such as building surface temperature, ground radiant heat, and airflow conditions, thereby enabling the estimation of thermal comfort indices, e.g., physiological equivalent temperature (PET), and building energy usage. This framework allows users to explore a variety of climate-resilient building surface strategies, e.g., green façades and cool paint applications, that improve thermal comfort while reducing wall heat gain and energy demand. By combining the autonomous reasoning capacity of LLMs with the rapid quantitative evaluation of lightweight physics-based models, the proposed system demonstrates potential for cross-disciplinary applications in sustainable urban design, indoor-outdoor environmental integration, and climate adaptation planning. The source code and data used in this study are available at: https://github.com/PgUpDn/urban-cooling-agent.

## 1 Introduction

Singapore's hot and humid climate and densely built-up urban areas can create uncomfortable outdoor conditions and high cooling demand indoors. For urban planners, the key questions are often practical: Where are the pedestrian heat-stress hotspots? Which buildings or street canyons are driving poor comfort and higher cooling load? Which surface strategies, such as cool coatings or green façades, offer the best trade-offs at neighborhood scale? Singapore is also investing in city-scale modelling initiatives to support such decisions. For example, Cooling Singapore 2.0 [1] aims to build a Digital Urban Climate Twin (DUCT) that links urban climate and building-energy models for testing cooling measures from island to neighborhood scale. However, running integrated comfort-energy studies at the neighborhood level is still laborious because workflows span multiple tools, many parameters, and time-consuming model preparation.

Recent work suggests that large language models (LLMs) can reduce this friction by acting as a "workflow layer" for simulation tasks. In building energy modelling, EPlus-LLM [2] shows how natural-language descriptions can be translated into EnergyPlus model inputs, lowering the manual burden of model setup. Other work [3] goes further by using an agentic workflow that generates model components and iteratively debugs them until an error-free EnergyPlus model runs. For indoor air quality and ventilation, an ontology-assisted GPT [4] framework integrates an LLM with CONTAM to support multizone airflow simulation and assessment. For thermal comfort and Heating, Ventilation, and Air Conditioning (HVAC) design tasks, recent studies have also explored LLM-assisted workflows [5] where designs proposed by models such as GPT-4o are later evaluated using Computational Fluid Dynamics (CFD) for thermal comfort and indoor air quality performance. Taken together, these studies show that LLMs can streamline simulation workflows, provided that results are tied to explicit inputs and solver outputs.

Building on this direction, we propose an agentic AI-enabled framework for thermal comfort and building energy assessment in tropical urban neighborhoods, motivated by Singapore's urban heat and cooling-demand challenges. The framework combines an LLM with lightweight physics-based models so that users can describe an urban design task in plain language, and the system can quickly produce quantitative outputs for both outdoor comfort (e.g., PET) and building cooling demand. Rather than replacing simulation, the LLM coordinates it:

---

* Corresponding author: lai_po-yen@a-star.edu.sg (Po-Yen Lai), yang_xinyu@a-star.edu.sg (Xinyu Yang)

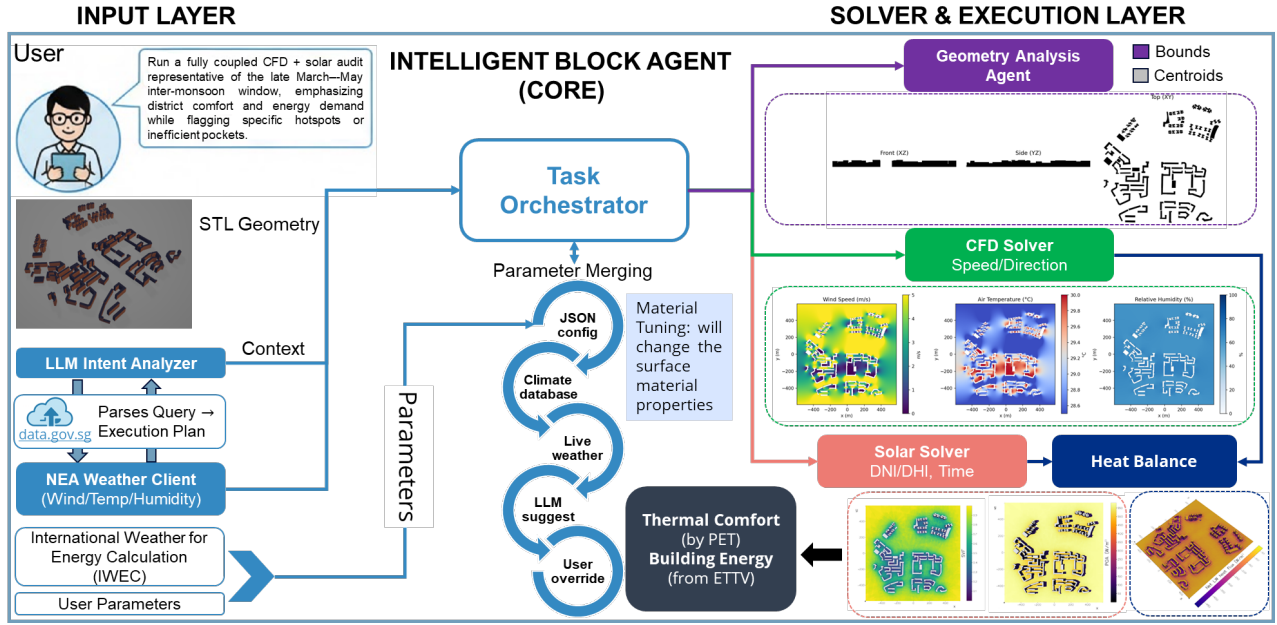

**Figure 1.** Agentic AI-Enabled framework for thermal comfort and building energy assessment in tropical urban neighborhoods.

it interprets intent, selects which analyses are needed (e.g., geometry, airflow, solar/radiation, comfort/energy), runs the appropriate solvers, and then produces a report that highlights hotspots and inefficient building clusters and points to the supporting outputs.

A central design feature is input governance and traceability. The LLM is advisory, but boundary conditions are controlled through a structured parameter object filled in a clear priority order: defaults from a JSON configuration, hourly weather from International Weather for Energy Calculation (IWEC) as the primary source of typical conditions, optional filling from Singapore meteorological station measurements only when needed, and LLM suggestions only when authoritative sources do not provide a value. Geometry is provided as Stereolithography (STL) files; the system automatically cleans meshes, assigns building IDs, and generates an index map so that reported hotspots can be traced back to specific buildings. A coupled airflow-solar run then produces maps and summary metrics (wind, temperature/humidity adjustments, radiation, surface temperature, PET, and cooling-load indicators), and the final report links every claim to file outputs and saved parameter snapshots. This enables rapid comparison of surface strategies (e.g., cool paints and green façades) while keeping the workflow reproducible and reviewable.

Overall, our contribution is an end-to-end, auditable workflow for tropical urban neighborhoods that connects outdoor comfort and building energy in one loop: intent → geometry → coupled airflow and solar simulation → comfort/energy metrics → readable report, with explicit logging of prompts, outputs, and parameter provenance → analysis and suggestion. This makes it practical to run and compare climate-resilient surface strategies quickly, while keeping the reasoning traceable enough for cross-disciplinary review and design discussion.

## 2 Methodology

The core orchestrator is implemented as a LangGraph-style state machine. At a high level, each run follows five stages (see **Fig. 1**):

1. **Intent analysis.** A GPT-5.1 model reads the user query and proposes which inputs and tools are needed (e.g., geometry, CFD solver, solar solver) together with initial parameter suggestions (e.g., weather).
2. **Geometry analysis.** STL files are loaded, cleaned, concatenated, and summarized. Building footprints are extracted and indexed for later cross-referencing with simulation outputs.
3. **Solver orchestration.** The Orchestrator Agent merges parameters from multiple sources, including configuration defaults (JSON config), climate databases (e.g., IWEC), real-time weather services, LLM advisories, and user overrides, according to a strict priority hierarchy. The merged parameters are then passed to the solver as its configuration and runtime inputs.
4. **Material strategy recommendation.** When surface material strategy recommendation is requested, the Orchestrator Agent plans a baseline-then-refined execution loop for the proposed material property adjustments (e.g., albedo, emissivity) and performs before/after comparison.
5. **Integration and reporting.** The LLM receives compact summaries of solver metrics (cooling load, PET, MRT, wind-comfort) and file paths, then drafts a human-readable analysis with actionable design recommendations. All prompts, responses and parameter snapshots are logged for full auditability.

**2.1 Agentic AI-Enabled Framework**

The agentic framework uses a single state object that is threaded through every stage of the LangGraph workflow. This object carries the user query, resolved simulation timestamp, LLM-suggested parameters, live weather readings, per-solver results, and error state, ensuring full traceability from input to output within a single run. Every LLM call across all agents is recorded by a shared LLMRecorder that writes (i) a structured JSON log with stage, prompt, response, and latency per call, and (ii) a human-readable verbose text log. Both are flushed after each interaction.

*2.1.1 Weather and Parameter Governance*

Solver parameters are resolved via a five-level priority chain. Following best practices from tool-using agents [6–8], we make the LLM advisory but not authoritative. The solver ultimately accepts a structured input object whose fields are populated according to the following priority order: (1) Configuration defaults: A JSON configuration file provides structural defaults for simulation parameters (mesh resolution, time stepping, material properties) (2) Climate database: The primary source of hourly weather conditions comes from standardized climate files (e.g., IWEC format), providing representative meteorological data for the target location. (3) Real-time weather service: An optional weather client can fetch measurements from public APIs, such as National Environment Agency (NEA) weather client for Singapore. These values only fill parameters that remain unset after the climate database pass. (4) LLM suggestions: intent analysis may propose context-aware refinements (e.g., adjusting wind speed for comfort-focused studies or selecting a representative timestamp for seasonal scenarios). LLM suggestions are applied only if both the climate database and real-time service left the field unset. (5) User overrides: Explicit user-supplied parameters have the highest priority and override all other sources. The final merged parameters are serialized to JSON files capturing not just the numeric values but also the provenance chain that produced them. This ensures full transparency and reproducibility.

*2.1.2 Geometry Analysis Agent*

The Geometry Analysis Agent loads all building STL files from the requested directory and performs validation and cleaning. Beyond concatenating individual STLs into a combined geometry file, the Geometry Analysis Agent: extracts XY bounding boxes for each building mesh; derives a unique identifier for each building based on file metadata; computes geometric statistics (height, footprint area, volume); renders a plan-view index map showing building outlines with ID labels at centroids. This index map serves as a lightweight reference that complements the more detailed CFD and solar visualizations. It answers practical questions such as *"which exact STL blocks correspond to this hotspot?"* by providing a consistent ID scheme across all outputs.

*2.1.3 CFD-Solar Solver Execution Agent*

For physics simulations, the Solver Execution Agent delegates to solver backends via wrapper classes. The wrapper prepares the output directory structure; resolves climate file paths; forwards the fully merged parameter set to the solver entry point. When radiation and CFD are both enabled, wind, temperature, humidity, and radiative fields evolve consistently over the simulated period (typically a representative 24-hour cycle). The solver produces a comprehensive suite of outputs used by the LLM to synthesize the final analysis report. These include VTK slices and surface fields characterizing atmospheric and thermal variables across spatial meshes. Additionally, the system generates visual renderings of simulation outputs, alongside a Metrics JSON file aggregating quantitative data such as cooling loads (kWh) and hotspot coordinates.

*2.1.4 Material Strategy Recommendation*

The Orchestrator Agent implements an automated baseline-then-refined workflow: *Step 1: Baseline Simulation*, to execute coupled CFD-solar simulation with default material properties, and then extract analysis metrics (e.g., per-building cooling loads, ground-level PET/MRT hotspots). *Step 2: Material Proposal*, to identify target buildings (highest cooling demand, proximity to thermal hotspots) and propose surface property adjustments based on heuristics and best practices. *Step 3: Comparative Analysis*, to rerun the simulation and compute delta metrics: cooling load reduction (kWh and %), peak demand changes. The Orchestrator Agent is also tasked with identifying buildings with the largest improvements and quantifying residual hotspots that require further interventions.

*2.1.5 Result Analysis and Recommendation Agent*

The final integration stage goes beyond simply reporting simulation outputs. The LLM is tasked to: (1) Synthesize multi-domain results: correlate CFD wind patterns with solar radiation exposure to explain observed thermal comfort patterns. (2) Identify design hotspots: rank locations by severity (PET, MRT) and attribute causes to specific geometric or material factors. (3) Quantify intervention impacts: report percentage improvements in cooling load and identify which building categories are most benefited. (4) Generate actionable recommendation: propose specific interventions prioritized by impact: (i) Material retrofits (cool roofs, reflective façades); (ii) Shading strategies (canopies, vegetation, louvers); (iii) Ventilation enhancements (opening orientations, wind corridors). (5) Reference output files: include paths to relevant VTK files, screenshots and metrics JSON.

## 2.2 Data Inputs

Before the agentic AI can invoke the lightweight physics-based models to assess urban microclimate conditions, two fundamental categories of input data must be prepared: comprehensive weather data and accurate geometric representations of the urban environment.

1. **Weather.** Simulations are driven by an hourly weather file (IWEC by default) providing, at minimum, wind speed/direction (10 m) and air temperature/relative humidity (2 m). For radiation, direct normal irradiance (DNI) and diffuse horizontal irradiance (DHI) are used when available; otherwise, they are derived from global irradiance via standard decomposition. In addition, geographic metadata (latitude/longitude/altitude) is required for solar position; Singapore–Changi is used as the default location for our work. User-specified meteorological overrides, when provided, supersede file values during time stepping.
2. **Geometry and Domain Setup.** Urban morphology is provided as a directory of STL meshes. The Geometry Analysis Agent validates, cleans, and concatenates them into a unified model. For the cases presented, the computational domain covers a 1×1 km urban area. The Geometry Analysis Agent auto-generates a ground plane expanded by a configurable buffer (default 1.2× the plan extents) to mitigate boundary effects. To capture micro-level airflow and thermal interactions around complex building morphologies, the spatial grid is discretized with an average horizontal resolution of 2 m.

These inputs establish the boundary conditions and computational domain necessary for simulation.

## 2.3 Lightweight Physics-Based Models

The framework invokes three model components: (1) CFD, (2) radiative heat transfer, and (3) thermal comfort and energy consumption assessment. The framework integrates these components to evaluate urban microclimate conditions and building thermal loads. Each model component can be independently called or combined depending on the user's analysis requirements and available input data.

### 2.3.1 CFD

The CFD component solves a 2D potential flow model to estimate wind velocity fields at multiple heights. Boundary conditions follow WMO standards: wind speed at 10 m reference height ($U_{10}$) and air temperature ($T_{air}$) and relative humidity (RH) at 2 m height.

Wind speed at arbitrary height $z$ is computed using the logarithmic wind profile law, $U(z) = U_{10} \cdot \frac{\ln(z/z_0)}{\ln(10/z_0)}$, where $U_{10}$ is the wind speed at 10 m reference height (m/s), $z$ is the height above ground (m), and $z_0$ is the aerodynamic roughness length set for the tropical urban environment [9]. This profile provides the free-stream velocity boundary condition for each horizontal slice in the computational domain.

To balance computational speed with spatial resolution, the airflow model employs a pseudo-3D approach. The solver executes 2D potential flow calculations at multiple discrete vertical slices (*e.g.*, at 2, 10, 20, 30, 40, 50, and 100m). The wind velocity vector for any given 3D building surface point is then derived via linear interpolation between these height-based slices. This allows the 2D solver to approximate vertical wind profiles and map local convective heat transfer coefficients onto the complex 3D building geometry.

Air temperature and humidity at each location are adjusted from the 2 m reference values based on local wind speed using an empirical tropical boundary layer model [10]. The temperature adjustment is computed as $T_{adj} = T_{air,2m} + \Delta T_{air}$, where the temperature change accounts for evaporative cooling, boundary layer mixing, radiative heating, and longwave cooling. Relative humidity is adjusted using Magnus-Tetens equation [11].

### 2.3.2 Radiation Model and Surface Energy Balance

The radiation model quantifies solar and thermal infrared exchanges on urban surface geometries. Surface temperatures evolve according to a time-dependent energy balance equation that couples radiative fluxes with thermal storage and convective heat transfer.

The surface temperature at each time step $t$ satisfies the transient energy balance [11]:

$$C_{face} \frac{\partial T_{surf}}{\partial t} = Q_{SW,in} + Q_{LW,in} - Q_{LW,emitted} - H(T_{surf} - T_{air}),$$

where $C_{face}$ is volumetric heat capacity (J/m³·K, default $0.5 \times 10^6$ J/m³·K), $T_{surf}$ is surface temperature (K), $Q_{SW,in}$ is incident shortwave radiation (W/m²), $Q_{LW,in}$ is incident longwave radiation (W/m²), $Q_{LW,emitted}$ is emitted longwave radiation (W/m²), $H$ is the convective heat transfer coefficient (W/m²·K), and $T_{air}$ is local air temperature (K).

After quantifying all radiation components, the Mean Radiant Temperature (MRT) is derived from radiative fluxes at the pedestrian level (typically 2 m) [11]. This calculation serves as a foundational step for evaluating advanced thermal comfort indices.

### 2.3.3 Thermal Comfort and Building Energy Evaluation

Thermal comfort is assessed using the Physiological Equivalent Temperature (PET) [13], which is derived from the Munich Energy-balance Model for Individuals (MEMI). This index integrates environmental parameters—including air temperature, mean radiant temperature, wind speed, and relative humidity—with personal factors such as metabolic rate, clothing insulation, and individual physical attributes.

Simultaneously, building cooling energy is estimated using a simplified 1D Conduction Transfer Function (CTF) model [14]. This calculates the conductive heat flux for each exterior mesh face based on the temperature difference between the computed exterior surface temperature and a fixed indoor setpoint (25°C). The total building cooling load is calculated by aggregating surface fluxes over the envelope. This is expressed as Energy Use Intensity (EUI) in cumulative kWh/m². Although this lightweight proxy bypasses the complexity of full-scale BEM engines like EnergyPlus, it maintains sufficient comparative accuracy to evaluate performance trade-offs and pinpoint energy-intensive façades.

## 3 Results and Analysis

The results presented here validate the proposed end-to-end, auditable workflow applied to a high-density tropical district. The analysis includes 2 scenarios corresponding to the established research scope: (1) A Baseline Geometric Audit (Diagnosis), where the Orchestrator Agent autonomously flagged spatiotemporal hotspots; and (2) A Closed-Loop Material Intervention (Mitigation Strategy), where the Orchestrator Agent proposed and verified surface strategies to mitigate heat stress.

Crucially, the results go beyond numerical output to demonstrate explicit traceability. The generated execution logs document the full provenance of simulation parameters, proving that the Orchestrator Agent's reasoning is accessible for cross-disciplinary review. The following sections detail how this transparent workflow allowed for the rapid comparison of climate-resilient strategies, revealing complex physical trade-offs, such as the interplay between envelope cooling loads and pedestrian radiative exposure that are essential for evidence-based urban design.

### 3.1 Scenario I: Autonomous Baseline Audit "From Semantic Prompt to Diagnostic Insight"

The first validation case tasked the Orchestrator Agent to perform geometry and microclimate modelling to evaluate outdoor thermal comfort and building cooling energy. We trace agent's autonomous execution path through 5 distinct stages, demonstrating its capacity to convert intent into actionable, physics-based design intelligence.

#### 3.1.1 Prompt (Intent Analysis)

We evaluate the Orchestrator Agent on two multi-building districts (denoted as District A and District B) in Singapore with Query: "*Run a fully coupled CFD + solar audit representative of the late March-May inter-monsoon*

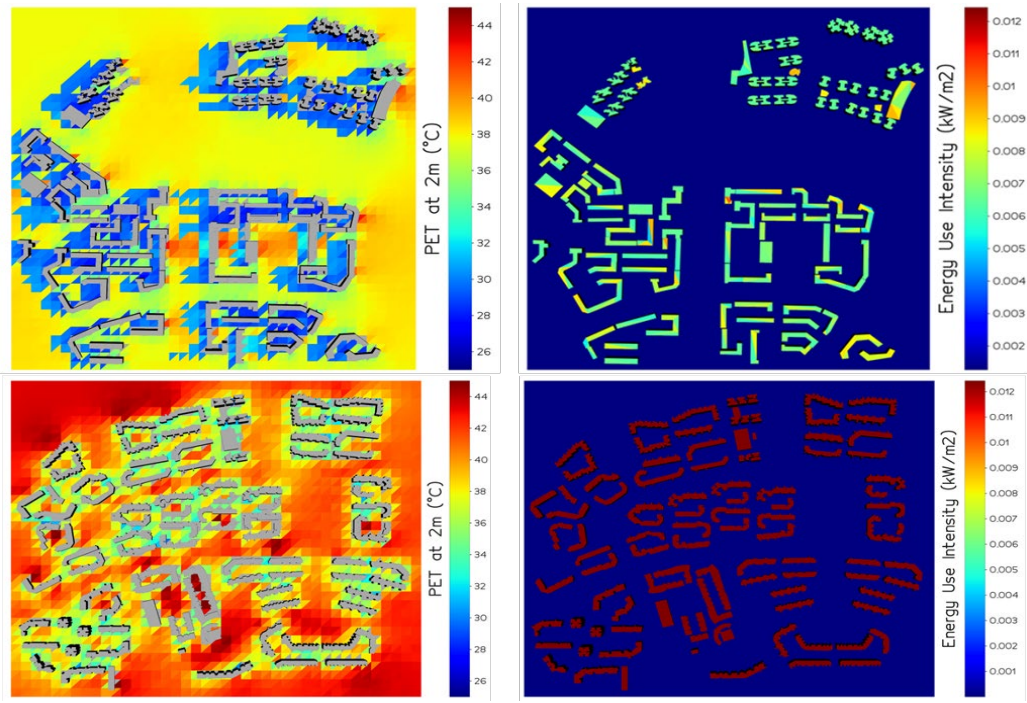

**Figure 2.** Pedestrian-level PET at 2 m (left) and building energy for cooling use intensity (right) for District A at 9 am (top) and District B at 3 pm (bottom), illustrating the key performance metrics of this study in terms of outdoor thermal comfort and building energy demand across the building cluster.

*window, emphasizing district comfort and energy demand while flagging specific hotspots or inefficient pockets.*" This prompt contained no explicit numerical parameters, demanding the Orchestrator Agent's autonomous interpretation of the climatic context.

*3.1.2 Input (Parameter Translation & Case Setup)*

For Baseline District A, the Orchestrator Agent autonomously interpreted "inter-monsoon" to select April 20 as the representative temporal baseline for its coupled CFD-Radiation simulation. Similarly, for Baseline District B, the Orchestrator Agent also configured its analysis for the same representative day. In both instances, the Geometry Analysis Agent instantiated a representative 1×1 km computational domain with an average grid resolution of 2 m to ensure adequate spatial fidelity around building clusters. The CFD solver was initialized with open boundary conditions (Dirichlet) to allow natural wind progression across the domain. The Orchestrator Agent autonomously configured for high-diffuse solar radiation (DNI ~110-205 W/m², DHI ~424-504 W/m²) and low-to-moderate wind velocities (2-6.2 m/s), consistent with the seasonal characteristics. Crucially, the Orchestrator Agent successfully set up the distinct underlying building geometries for each run, demonstrating its flexibility in handling physical inputs.

*3.1.3 Major Output (Key Performance Metrics)*

For both baseline configurations, the Orchestrator Agent executed coupled CFD and radiative transfer solvers. The analysis prioritized two primary output streams (see **Fig. 2** as an example):
1. **Outdoor Comfort.** Spatially resolved hourly PET maps at the pedestrian level (2m above ground).
2. **Building Energy.** Hourly and cumulative Cooling Energy Demand calculated for each building envelope based on convective and radiative heat flux.

*3.1.4 Analysis (Anomaly Detection Across Variants)*

Based on the spatial configurations for District A and District B illustrated in **Fig. 3**, the Geometry Analysis Agent utilized the labelled building IDs. A localized analysis was conducted to identify the following:
1. **Thermal Stress Hotspots (PET).** District A: Detected an absolute PET peak of 52.18°C at 1:00 PM in a plaza near buildings b025 and b057. District B: Identified a similarly extreme PET peak of 52.35°C at 1:00 PM in a different plaza near buildings b040 and b055. This demonstrated the Orchestrator Agent's robustness in consistently pinpointing specific coordinates of extreme thermal stress (>50°C), regardless of the urban layout.
2. **Energy Intensity Profiling (Cooling Load).** District A: Flagged high-rise structures (e.g., b078, b070–b077) as energy outliers, with normalized cooling loads much higher than others. District B: Similarly identified buildings (e.g., b086, b088) with comparable high normalized loads. By normalizing total consumption against envelope area, the Orchestrator Agent successfully decoupled absolute size from thermal inefficiency, isolating the most energy-intensive typologies in both scenarios.

*3.1.5 Summary and Strategic Recommendations*

For each baseline run, the Recommendation Agent synthesized these metric-specific findings into coherent diagnostic reports and formulated tailored design interventions. It provided accurate physical reasoning, explicitly linking the high PET values to specific combinations of solar exposure (e.g., high sky view factor) and low wind speed (e.g., stagnant airflow), and attributing high cooling loads to unshaded façade orientations. Completing the loop, the Recommendation Agent autonomously formulated recommendations based on the specific building location (see **Fig. 3**). For instance, in District A, it suggested overhead shading for the b025-b057 zone and façade optimization for high-load blocks. In District B, it proposed similar interventions, including shading for the b040/b055 plaza and specific recommendations for the most energy-intensive buildings like b086 and b088. These targeted proposals confirm the Recommendation Agent's capability to translate quantitative PET and energy data into actionable, location-specific design strategies that align with identified physical causes.

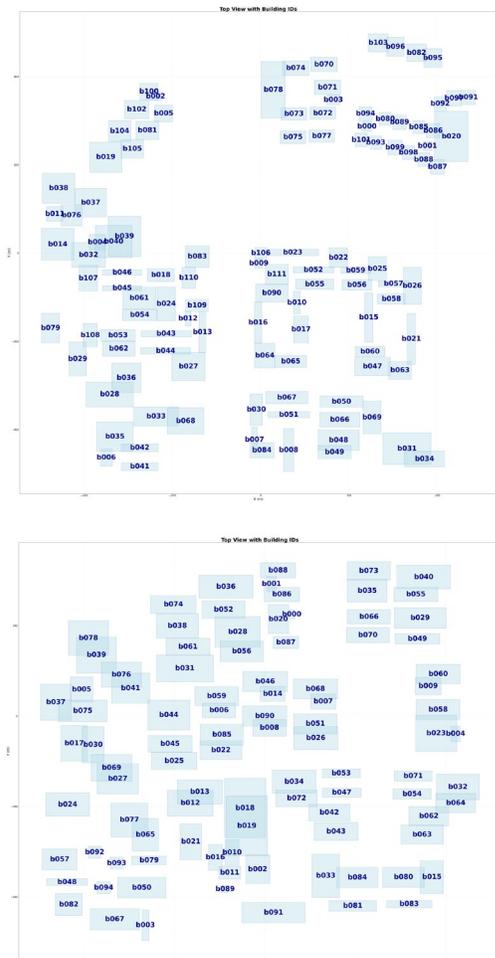

**Figure 3.** Top-down plan views of building layouts and ID assignments for District A (top) and B (bottom). The labelled building IDs enable a localized analysis of simulation results, specifically for identifying critical thermal hotspots.

### 3.2 Scenario II: A Closed-Loop Material Intervention and the Discovery of the "Albedo Penalty"

In District A, the second test case evaluates the Orchestrator Agent's autonomous intervention via a closed-loop hypothesis-verification cycle. Rather than using resource-intensive numerical optimization, the Orchestrator Agent employs an iterative, heuristic-based inference loop. It identifies high-gain surfaces, applies ASHRAE-standard high-albedo materials [14], and verifies improvements through targeted "baseline-then-mitigation" simulations. This demonstrates the Orchestrator Agent's ability to logically navigate design options rather than relying on brute-force mathematical convergence.

*3.2.1 Autonomous Hypothesis and Parameter Overrides*

1. Prompt: The Orchestrator Agent received a new directive to "*... propose and apply surface material changes (albedo/emissivity) to improve thermal comfort and reduce cooling load.*" This directly followed the Baseline Audit.
2. Input: The Orchestrator Agent chose April 15 as a representative inter-monsoon day and the original building distribution from Scenario I (Baseline Audit). The Recommendation Agent autonomously generated a modification plan. It then executed a targeted parameter override within the simulation environment, setting the albedo of roofs, walls and ground surfaces with higher reflectance values, while keeping emissivity fixed at ~0.9. This constituted the new set of physical properties for the re-simulation, ensuring a direct comparative analysis against the baseline.

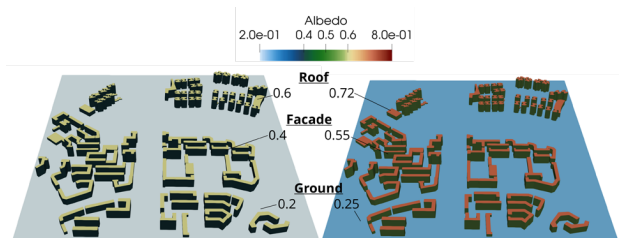

**Figure 4.** Spatial distribution of prescribed albedo values for roofs, façades, and ground surfaces in the study area (District A), comparing the baseline configuration (left) with the LLM-suggested configuration (right).

*3.2.2 Comparative Analysis (Energy-Comfort Trade-Off)*

The Orchestrator Agent executed a new simulation and performed a differential analysis between the Baseline and the mitigated scenarios (see **Fig. 4**). The results validated the effectiveness of the intervention for building performance but revealed a critical physical divergence for outdoor comfort:

1. **Building Energy Reduction.** The Recommendation Agent successfully verified that the high-albedo strategy reduced envelope heat gains. For instance, peak cooling power on representative sun-exposed façades dropped measurably (e.g., from ~784 W to ~740 W per element at noon), confirming the hypothesis for energy efficiency.
2. **The "Albedo Penalty" on Comfort.** Conversely, the Recommendation Agent detected that pedestrian comfort did not improve. In fact, at the identified noon hotspot, the PET value slightly increased ~1°C. The Recommendation Agent correctly isolated the physical cause in its comparison report: while surface temperatures decreased (reducing longwave emission), the reflected shortwave radiation from the brighter ground significantly increased the mean radiant load on pedestrians.

**Table 1** summarizes the comparative results for the baseline and mitigated configurations across the prioritized high-load buildings.

**Table 1**. Percentage reduction in daily cooling energy loads for prioritized high-load building envelopes in District A following mitigation strategy implementation.

| Building ID | Description | Reduction |
|---|---|---|
| b091 | Tall corner tower | -10.70% |
| b090 | Tall central tower | -10.10% |
| b085 | 46 m tower | -9.60% |
| b076 | 37.7 m block (NW) | -9.70% |
| b077 | 37.7 m block (S) | -11.40% |
| b044 | 34.9 m high-load block | -8.50% |
| b040 | 32.1 m block (NE) | -10.80% |
| b036 | 32.1 m block | -9.00% |

*3.2.3 Strategic Synthesis and Refined Recommendations*

Recognizing this non-linear trade-off, the Recommendation Agent refined its design advice in the final report. It moved away from a blanket "cool surface" recommendation, instead proposing a decoupled strategy: prioritizing high-albedo coatings for roofs (to lower cooling demand) while advising against high-reflectivity pavements in sun-exposed plazas to avoid exacerbating radiative stress on pedestrians. This ability to self-correct based on simulation feedback demonstrates the Orchestrator Agent's potential to navigate complex, conflicting urban physics objectives without human supervision.

## 4 Conclusion and Future Work

This study validated an autonomous framework for urban environmental analysis through two scenarios. Baseline Audit Scenario demonstrates the Orchestrator Agent's diagnostic precision in translating semantic intents into rigorous boundary conditions to identify thermal and energy hotspots. Closed-Loop Intervention Scenario proves its ability to execute hypothesis-driven recommendation, correctly simulating the counter-intuitive "albedo penalty" and navigating non-linear physical trade-offs without human supervision. From an agentic system perspective, these experiments illustrate how an LLM can sit "on top of" a physics engine without overriding its domain logic. The LLM handles intent analysis and explanation, while trusted weather sources maintain strict control over physical boundary conditions. This architecture ensures the workflow is conversational, yet scientifically deterministic. Compared with human-in-the-loop configuration or fixed pipelines, the agentic LLM orchestration enables faster setup, less specialist dependence, and higher reuse across audit and intervention tasks.

Beyond these benefits, the Execution Agent is designed to be backend-agnostic. The wrapper interface is backend-agnostic and can also dispatch to full-fidelity solvers or AI surrogate models. This modularity enables scaling from quick exploratory analyses to high-resolution design validation without changing the agentic workflow. Future extensions include (1) Scalability: Implementing memory-aware meshing and tiling to handle large-scale geometry collections. (2) Standardization: Automatically computing standardized KPIs (e.g., Lawson/NEN wind criteria, PET thresholds) from output data to enable direct regulatory compliance checking. (3) Interaction: Developing an interactive dashboard to expose logs and metrics in real-time. This "report-as-you-simulate" approach would allow non-experts to interrogate data and trace parameter provenance, building trust in the agentic recommendations.

## Generative AI Disclosure

A generative AI tool was used during the preparation of this manuscript to improve clarity and grammar. All technical content, claims, and conclusions were verified by the authors.

## Acknowledgement

This research was supported by the National Research Foundation, Singapore through the AI Singapore Programme, under the project "AI-based urban cooling technology development" (Award No. AISG3-TC-2024-014-SGKR).